\documentclass[conference]{IEEEtran}

\usepackage{graphicx}
\usepackage{booktabs}
\usepackage{array}
\usepackage{subcaption}
\usepackage{url}
\usepackage{listings}
\usepackage{color}
\usepackage{multirow}
\usepackage{paralist} 
\usepackage{amsmath}
\usepackage{amssymb}
\usepackage{enumitem}
\usepackage{booktabs}
\usepackage{float}

\usepackage[ruled,vlined]{algorithm2e}

\newtheorem{definition}{Definition}
\newtheorem{theorem}{Theorem}
\newtheorem{observation}{Observation}
\newtheorem{lemma}[theorem]{Lemma}
\newenvironment{pfof}{\noindent{\em Proof:} }{ \hfill $\blacksquare$\\ }

\newcommand{\MECH}{CSOPT}
\newcommand{\hc}{\mathbf{\hat{c}}}
\newcommand{\chT}{\mathcal{\hat{T}}}
\newcommand{\cT}{\mathcal{T}}

\usepackage{listings}

\definecolor{dkgreen}{rgb}{0,0.6,0}
\definecolor{gray}{rgb}{0.5,0.5,0.5}
\definecolor{mauve}{rgb}{0.58,0,0.82}

\begin{document}
\title{Privacy Preserving and Cost Optimal Mobile \\ 
Crowdsensing using Smart Contracts on Blockchain}
\author{
\IEEEauthorblockN{Dimitris Chatzopoulos$^{*}$, Sujit Gujar$^{\&}$, Boi Faltings$^{\$}$, and Pan Hui$^{*}$$^{\#}$}
dcab@cse.ust.hk, sujit.gujar@iiit.ac.in, boi.faltings@epfl.ch, panhui@cse.ust.hk \\
  $^{*}$HKUST, $^{\&}$IIIT Hyderabad, $^{\$}$EPFL, $^{\#}$University of Helsinki
}

\maketitle

\begin{abstract}
The popularity and applicability of mobile crowdsensing applications are continuously increasing due to the widespread of mobile devices and their sensing and processing capabilities. However, we need to offer appropriate incentives to the mobile users who contribute their resources and preserve their privacy. Blockchain technologies enable semi-anonymous multi-party interactions and can be utilized in crowdsensing applications to maintain the privacy of the mobile users while ensuring first-rate crowdsensed data. In this work, we propose to use blockchain technologies and smart contracts to orchestrate the interactions between mobile crowdsensing providers and mobile users for the case of spatial crowdsensing, where mobile users need to be at specific locations to perform the tasks. Smart contracts, by operating as processes that are executed on the blockchain, are used to preserve users' privacy and make payments. Furthermore, for the assignment of the crowdsensing tasks to the mobile users, we design a truthful, cost-optimal auction that minimizes the payments from the crowdsensing providers to the mobile users. Extensive experimental results show that the proposed privacy preserving auction outperforms state-of-the-art proposals regarding cost by ten times for high numbers of mobile users and tasks.
\end{abstract}

\section{Introduction}

The wide dissemination of smartphones that are programmable and employed with sensors gave birth to \emph{crowdsensing} applications such as environment monitoring, mobile social recommendations, public safety and others. 
Mobile crowdsensing is a paradigm that utilizes the ubiquitousness of the mobile users who are carrying smartphones and can collect and process data. 
\textit{Crowsensing Service Providers} (CSPs) request sensing \textit{tasks} to \textit{mobile users} ($MU$s) who deliver these tasks in order to get paid. 
Crowdsensing tasks can be categorized based on characteristics inherent to the tasks or the participants\footnote{We are using the terms "mobile users" and "participants" interchangeably and depending on the context.}. 
Two usual dimensions are event based vs. continuous, and spatial vs. non-spatial. These dimensions are independent of each other, and any combination is possible. 

In this work, we focus on \textit{event-based spatial crowdsensing tasks} that are associated with geographic locations where the mobile users perform them~\cite{Ganti:2010:GPS:1814433.1814450, yan2011crowdpark}. The challenges are two-fold: \textit{(i)} the mobile users are sensitive about the secrecy of their locations and may not participate to avoid any leakage. Also, they may even try to spoof their locations to avoid the cost of moving the required locations. \textit{(ii)} A second challenge is the calculation of the payments to $MU$s for their participation. The \textit{participation cost} of each user is private information and depends on several factors. As a consequence, mobile users are motivated to misreport their actual costs to obtain higher payment, and hence incentives are needed. Truthful auctions are designed in such a way to force participants to report their true participation cost. This feature enables optimal task assignment to the participants in such a way to minimize the payments to the employed mobile users~\cite{nisan07chap9}. 

\begin{figure*}
\centering
\includegraphics[width=1.89\columnwidth]{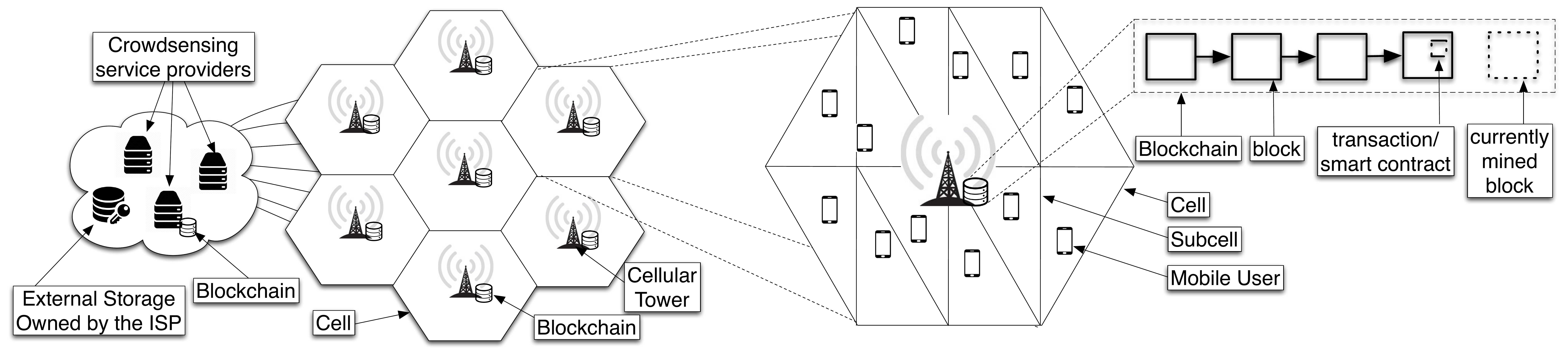}
\caption{The examined ecosystem. A blockchain is maintained by \textit{(i)} the ISP and \textit{(ii)} anyone else interested in the stored data. Smart contracts are used to coordinate the interactions between the ISP, the CSPs and the mobile users.}
\label{fig:arch}
\end{figure*}

We consider participants who are not willing to reveal their identities and locations regardless of the number of the tasks they have delivered. Although Internet service providers (\textit{ISPs}) are aware of users' identities and locations, they are not allowed to reveal them to third-parties~\cite{EUrights}. We propose to use the capabilities of ISPs supplemented by smart contracts over \emph{blockchains} to design a system for privacy-preserving crowdsensing that minimizes CSPs' cost. We propose a model where CSPs send crowdsensing requests to an ISP who transforms them into tasks and runs a cost-optimal auction to the suitable cells to allow the $MU$s on these cells to express their interest in the tasks via truthful bidding. The ISP is assisted by a blockchain, similar to Ethereum~\cite{wood2014ethereum} and Hawk~\cite{7546538} or Hyperledger Fabric~\cite{cachin2016architecture}. To build such crowdsensing system, we address the following questions:
\begin{itemize}
\item[$\mathcal{Q}_1$:] How to ensure a CSP that the data has been submitted by users at the indicated locations?
\item[$\mathcal{Q}_2$:] How to preserve the privacy of mobile users from CSPs, even if they have submitted location-specific data?
\item[$\mathcal{Q}_3$:] How to assign crowdsensing tasks to mobile users who are interested in subsets of tasks in a cost-optimal way and incentivize them to report their costs truthfully?
\end{itemize}
For $\mathcal{Q}_1$ and $\mathcal{Q}_2$, we leverage the confidentiality assurance from ISPs. ISPs guarantee the execution of CSPs' tasks at the desired locations. To build such trust across CSPs, ISPs, and $MU$s, we use a blockchain and \emph{smart contracts}. To address $\mathcal{Q}_3$ we design an auction using game theory. 

\noindent\textbf{Why blockchain?} \emph{Blockchain} is a distributed mechanism that stores data in the form of transactions and can offer additional functionalities such as \textit{transactional privacy} and \textit{smart contracts}. It is maintained by interconnected nodes that are responsible for securing the network, and keeping everyone in the system in sync. Anyone interested in maintaining a blockchain, and, as a consequence, in having access to the stored data can partake. Blockchains have been used in mobile environments such as for automated payments between mobile devices in cooperative application execution scenarios~\cite{8024034} and for enabling small payments between mobile users in environments without internet connectivity~\cite{Chatzopoulos:2016:LAP:2942358.2947401}.

In our scenario, we use the cellular access points of the ISP network to maintain a blockchain, but we assume that anyone (e.g., the CSPs) can participate. Transactional privacy guarantees that the identity of the creator of one transaction cannot be revealed. This functionality is used to hide users' identities. Smart contracts are software processes that are executed whenever a transaction is calling them when it is added to the blockchain. Ethereum allows any application to be deployed, using smart contracts, on the blockchain~\cite{wood2014ethereum, buterin2014next}. For a smart contract to be executed, a certain amount of credits has to be transferred to their address. We use this feature to enforce payments. Blockchains are more preferable to servers for various reasons. First of all, they are open and append-only mechanisms that can guarantee that the stored data can not be modified. This feature guarantees the integrity of the stored data. Second, the use of the smart contracts allows anyone to examine the validity of the produced outcomes~\cite{husearching}. 

Figure~\ref{fig:arch} shows the examined architecture and the participating entities (CSPs, ISP, $MU$s). Cellular towers can estimate, with high accuracy, the current location of each user and for that reason, we assume that a cell can be further split into smaller areas to allow the submission of crowdsensing requests with high granularity.  
The ISP employs smart contracts to \textit{(i)} give access to CSPs to the collected data they requested, \textit{(ii)} preserve the privacy of mobile users, \textit{(iii)} run auctions, \textit{(iv)} pay mobile users and \textit{(v)} get paid by the CSPs. This means that the trinity of CSPs, mobile users and the ISP interact with each other using smart contracts that are stored and executed in the blockchain. 
In summary, our contributions are the following:

\textbf{Contributions}: We address the problem of privacy preserving crowdsensing in a cost-optimal way by proposing the use of an ISP as the intermediary between CSPs and mobile users. ISP uses smart contracts over a blockchain to preserve the privacy of mobile users while ensuring the validity of their locations. As far as incentives for mobile user participation, we have designed a truthful, computationally efficient auction, called \MECH. The cost-effectiveness of \MECH\ is compared with a state-of-the-art algorithm, and the performance of the proposed smart contracts is depicted using Ethereum.

\section{Related Work}
\label{sec:rel_work}
Mobile users are motivated to spoof their location to preserve their privacy and potentially decrease their execution cost~\cite{940014, Tippenhauer:2011:RSG:2046707.2046719, 5168926}. Privacy concerns might even discourage users from participating. Depending on the type of a task, the potential privacy breach changes. For example, a task that requires an $MU$ to report the time needed to travel from one location to another by traveling at the time of the request, might lead to the disclosure of their current location and potentially sensitive addresses or even their identity through location-based attacks~\cite{Pournajaf:2016:PPM:2935694.2935700}. In the case of frequent participation, even if participants are using pseudonyms, their trajectory might reveal their sensitive locations or commutes~\cite{krumm2007inference} and even eventually disclose their identities~\cite{gambs2014anonymization}. 

Although there is high research activity on mobile crowdsensing, neither blockchain nor smart contracts have been used in the existing proposals, to the best of our knowledge. Proposed crowdsensing architectures are composed of a mobile application and a server that is responsible for the collection and processing of the sensed data. Localized analytics on the mobile devices are often performed to preserve users' privacy and reduce the amount of the data sent to the server~\cite{ganti2011mobile}. Furthermore, similar to the deployment of smart contracts in the orchestration of the crowdsensing process, the authors of \cite{Ra12a} develop Medusa, a framework to develop crowdsensing applications. However, the authors consider a crowdsensing application provider that is using cloud resources and do not provide any privacy guarantees to the mobile users. Similarly to this work the authors of~\cite{Merlino2016623} propose the ungearing of the crowdsensing provider from the physical resources that are responsible for the data gathering and processing. However they consider cloud infrastructure providers for that role, who do not provide any privacy guarantees. Liu \textit{et. al.}~\cite{5984882} consider the employment of a network provider to handle the crowdsensing process but they do not consider an auction in the determination of the users' cost since they assume that the ISP will determine the credits each $MU$ gets.

In our proposal the $MU$s are paid based on their costs and for which we rely on auctions. A \emph{cost optimal auction} is an auction that minimizes the expected payments of the CSP subject to feasibility constraints~\cite{Myerson81}. In his seminal work, Myerson~\cite{Myerson81} introduces the notion of optimal auction and designs one for selling a single unit of a single item. Our case is multiple units of multiple items (homogeneous but location specific tasks and hence we refer to it as multiple items). In economic terms, it falls under the category of \emph{multi-unit combinatorial auctions}, which is in general hard to solve. Optimal multiple items auctions have been proposed for specific settings. For example, Cai \emph{et. al.}~\cite{Cai12} consider additive value settings.  Iyengar and Kumar \cite{IYENGAR08} design an optimal multi-unit but single item auction. Mechanism design theory has been used for crowdsensing to design incentives~\cite{koutsopoulos13,Yang16,Zhao14auction}.  Koutsopoulos~\cite{koutsopoulos13} designs an optimal auction for crowdsensing. However, there is no deadline or no limit on the amount of the work a participant is willing to do or any location specific tasks. Hence his work is single item multiple units.  Karaliopoulos \emph{et.al.}~\cite{Karaliopoulos15} and Yang \emph{et. al.}~\cite{Yang16} consider a setting the same as ours except for the fact that we offer the flexibility to the ISP to assign $MU$s a subset of tasks instead of a complete set of the tasks in which they show interest. This leads to cost saving to the CSP as we do not repeat any task more than required. In \cite{Karaliopoulos15} the authors design approximate cost minimizing solutions, but do not consider the strategic behaviour of the participants. Yang \emph{et. al.} consider designing a truthful auction for the settings very similar to ours. However, their goal is to design a computationally efficient and truthful auction. 
In our settings, we allow ISP to allocate an $MU$ any subset of set of tasks in which it has shown an interest. In addition, we minimize the total expected payment made by the CSP. 
Another approach to offer incentives is fixed rewards rather than auction based mechanisms. For example. the incentive schemes proposed in \cite{Goel14,Bhattacharya10,Radanvoic16b,Radanovic16c}. However, in such settings the $MU$s are either overpaid 
or there is a need for more $MU$s, since the payments are less than their actual cost of delivering the task. For more on game theoretic approaches on incentive design, the readers are referred to~\cite{nisan07chap9}.


\section{Mobile Crowdsensing using Blockchain}\label{sec:main}
CSPs send their requests to the ISP who uses a smart contract to register the requests and collect the fees from the CSP for their requests. Then the ISP runs the auction using another smart contract to provide transparency in the selection of the proper mobile users. This smart contract forces the $MU$s to pay a participation fee that they will lose if they are selected and not submitted their measurements. Before the auction, the ISP creates a temporary id for each user in order to preserve the identity of the MUs. Next, the ISP uses another smart contract to collect participation proofs from the $MU$s and pay them. The $MU$s will only submit their collected data to the ISP but they will create a transaction that includes a hash of their data in order to trigger the smart contracts that pays them. Also, a fourth smart contract will give access to the CSP to the collected data. In order to execute this smart contract and get access to the collected data, the CSP has to transfer as many credits as the auction cost. The proposed smart contracts can be managed via mechanisms similar to~\cite{Hu:2018:HIE:3211933.3211935}. Before going into the details of our proposal, we introduce the used notation.

\subsection{Notation and Assumptions}
We consider a set of mobile users ($MU$s), $\mathcal{N}$, of size $|\mathcal{N}| = n$, one crowdsensing service provider, CSP, and one Internet service provider, ISP (the model can be generalized for more than one CSPs). Whenever the CSP sends a request, $CS_{req}$, to the ISP with deadline $D$, the ISP maps the request to a set of tasks $\mathcal{T}$ and runs an auction on the appropriate cells. Each cell $\mathcal{Z}_i \in \mathcal{Z}$ is further split into areas $z_{ij} \in \mathcal{Z}_i$. Each mobile user $MU_i$ is associated with a location, $l_i = z_{jl} \in \mathcal{Z}_j \in \mathcal{Z}$ and is able to bid for the set of tasks $\mathcal{T}_i \subset \mathcal{T} = \{T_{i1},T_{i2},\ldots,T_{ik_i}\}$ that it can deliver based on its current location and using the proper sensor before~$D$. Each $MU$ successfully completes a task with probability $\alpha$. The CSP requires enough $MU$s at each location, in order for the probability to successfully receive the task to be at least $\beta$. Given that the mobile users need to move to the appropriate locations to do the tasks, we assume that the maximum number of tasks a user can do is $k$. The cost for the execution of the first task for $MU_i$ is $c_{i1}$, for the second task $c_{i2}$ and so on. We denote its cost vector by $\mathbf{c}_i \in \mathbf{C}_i$ and private information as $\theta_i=(\mathbf{c}_i,\cT_i)$, which is called its \emph{type} in mechanism design theory. It submits a bid ${b}_i = (\hc_i,\mathcal{\hat{T}}_i)$, where $\hc_i$ is its reported cost and $\chT_i \subset \cT_i$ the reported tasks of interest. The ISP collects all the bids $\mathbf{b}=(b_1,b_2,\ldots,{b}_n)=({b}_i,{b}_{-i})$ where ${b}_{-i}$ represents the bids from all $MU$s except $MU_i$. 
Upon receiving $\mathbf{b}$, ISP determines the assignments, $\mathcal{A}(\mathbf{b})=(\mathcal{AT}_1,\mathcal{AT}_2,\ldots, \mathcal{AT}_n)$, where $\mathcal{AT}_i \subset \chT_i$ is a set of tasks assigned to $MU_i$, and the payments $\mathbf{p}(\mathbf{b})=(p_1(\mathbf{b}),p_2(\mathbf{b}),\ldots,p_n(\mathbf{b}))$. 
Then, the CSP is informed about the availability of the requested data. 

Let $n_i = \mid\mathcal{AT}_i\mid$ denote the number of the tasks that assigned to $MU_i$. With these, $MU_i$ obtains utility $u_i(\cdot)$ by participating in the crowdsensing auction. For given bids $\mathbf{b}$ and true type $\theta_i$, $u_i$ is given by:
\begin{equation*}
u_i(\mathbf{b};\theta_i) = p_i(\mathbf{b}) - \sum_{j=1}^{j=n_i} c_{ij}.
\end{equation*}
We drop argument $\mathbf{b}$ and just use $p_i, n_i$ whenever it is clear from the context.  We use either $b_i$ or  $(\hc_i,\chT_i)$ based on convenience in the proof and the same for $\theta_i$ and $(\mathbf{c}_i,\cT_i)$. In our model, we assume that the $MU$s will not submit their bids for the tasks they cannot do. This is a valid assumption and we show how to ensure this using smart contracts. We also assume that there is enough competition between $MU$s and even if we exclude one $MU$, the request can still be served. In the next section, we explain the role of the blockchain in our model and after that, the required smart contracts in order for the model to be functional.

\subsection{The use of Blockchain}\label{sec:blockchain}

\noindent There exist two types of interactions in our model: 

	\subsubsection{\textbf{Conventional}} There are three interactions of this type. \textit{(i)}~The requests from the CSPs that contains the characteristics of the tasks ($SC_{req}$), \textit{(ii)} the advertisement of the tasks from the ISP to the $MU$s and the initiation of the auctions \textit{ADV}($\mathcal{T},\mathcal{C}$), and \textit{(iii)} the submission of the sensed data from the mobile users.
    \subsubsection{\textbf{Blockchain-based}} These interactions take the form of transactions and are stored in the blockchain. Such interactions require the interacting entities to have an account. \emph{Transactions} are the building blocks of blockchains, represent interactions between two or more entities and are associated with some data. In its simplest form, a transaction represents the exchange of money~\cite{nakamoto2012bitcoin,7423672} but it can also be used in more complicated forms, like the one where a mobile user submits a sensor reading. There are two types of accounts, the externally owned ones (CSPs, $MU$s) and the smart contracts. Smart contracts are special types of accounts, which have a set of functionalities, are stored on the blockchain, and are uniquely identifiable. They also have their own storage, which can be changed whenever they are triggered by a transaction. Smart contracts allow us to have general purpose computations on the chain. Whenever such transactions are created, every miner automatically executes the contract and considers the data included in the transaction as an input. Then, the whole blockchain network operates as a distributed virtual machine. All the remaining interactions belong to this type. 

\subsection{Proposed Smart Contracts}\label{sec:CSPandISP}

Whenever the ISP receives a $CS_{req}$, it creates a transaction which is signed with the public key of the CSP. The transaction includes the timestamp of the request, the deadline $D$ and the address of the smart contract called \textit{Request Registration} (RR) that the CSP will call after the deadline in order to get access to the collected data on the external database of the ISP. Before the deadline, the ISP creates another smart contract, called \textit{Data Access} (DA), and stores its address to RR. DA contains the credentials to the external database where the ISP stores the collected data. The credentials are encrypted using the public key of the CSP in order to allow only the CSP that submitted the request to get access to the collected data. When the CSP, will trigger RR, it has to create a transaction with the RR as the destination and in order for the smart contract to be executed, the CSP has to include enough credits (in the Ethereum project, these credits are called ether~\cite{wood2014ethereum}). In this way, the CSPs have to pay a fee included by the ISP in order to get the address of DA. Then for the execution of DA, the CSP will have to pay the amount the ISP paid to the mobile users after the collection of the data. The ISP is responsible to store in RR the address of DA and the hash of the collected data for $CS_{req}$. If these two entries are not filled before the deadline, the RR generates a transaction from the ISP to the CSP and transfers back the credits. 

\begin{table}[t]
\begin{center}
\begin{tabular}{l l}
\toprule
Name & Type \\
\midrule
$SC_{req}$	& Conventional \\
\cmidrule(l){1-2}
Request Registration (RR)	& Blockchain-based \\ 
\cmidrule(l){1-2}
Data Access (DA)			& Blockchain-based \\
\cmidrule(l){1-2}
\textit{ADV}($\mathcal{T},\mathcal{C}$) & Conventional \\
\cmidrule(l){1-2}
Crowdsensing Optimal (CSOPT) &  Blockchain-based \\
\cmidrule(l){1-2}
Submission of Sensed Data &  Conventional \\
\cmidrule(l){1-2}
Mobile User Payment (MUP) &  Blockchain-based \\
\bottomrule
\end{tabular}
\end{center}
\caption{List of possible interactions among the entities. For conventional interactions the ISP employs a server that receives the requests from the CSPs. 
}
\label{tab:interactions}
\end{table}

\begin{figure}[t]
\centering
\includegraphics[width=0.9\columnwidth]{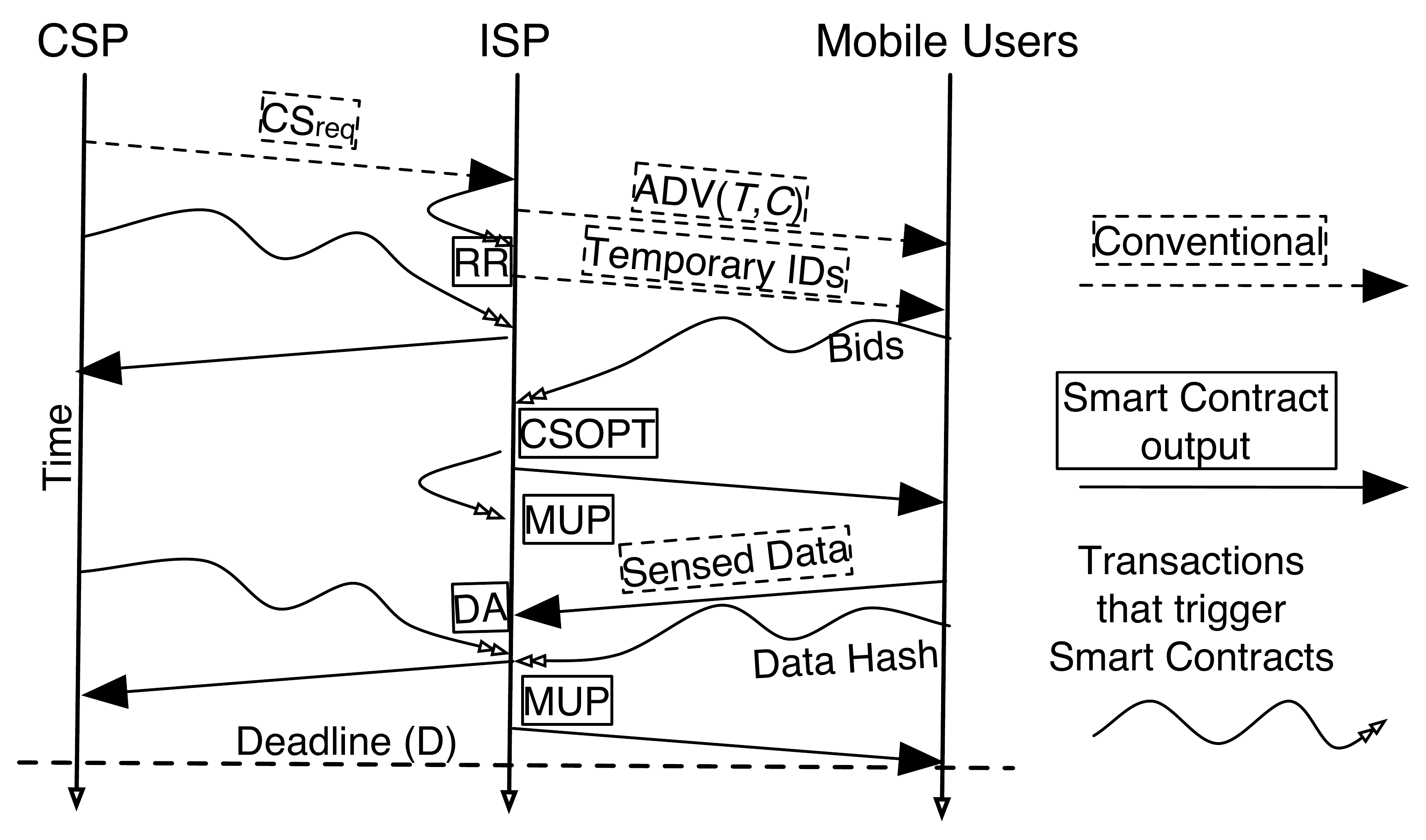}
\caption{Interactions between the crowdsensing service provider, the Internet service provider and the mobile users.}
\label{fig:proto}
\end{figure}

The ISP, after the reception of $CS_{req}$, decides which are the locations of interest and broadcasts the characteristics of the tasks to the $MU$s on these locations~(\textit{ADV}($\mathcal{T},\mathcal{C}$)). Also, the ISP 
creates a temporary account in the blockchain for each of the $MU$s that it will be used on for this auction. Each mobile user, $MU_i$, submits a bid ${b}_i = (\hc_i,\mathcal{\hat{T}}_i)$ in a form of a transaction, to express its interest on executing tasks $\mathcal{\hat{T}}_i \in \mathcal{T}$, to the blockchain using its temporary address. All the bids are submitted to the designed smart contract called \textit{Crowdsensing Optimal} (CSOPT) that produces a new transaction that contains the task assignment. The optimality of CSOPT is presented in Section \ref{sec:ISPandMUS}.  In this way, the ISP is not able to manipulate the bids, the CSP is also able to verify the cost of its request and the mobile users are not revealing their identity.
Since each $MU$ needs to transfer certain credits in order to trigger \MECH, \MECH\ after the production of the assignment creates a transaction and sends back to the non selected $MU$s the credits they spent for the auction. The selected ones will get their credits back after the completion of their tasks. If they fail to submit their tasks, they will lose their credits.  Also, \MECH\ triggers another contract called \textit{Mobile User Payment} (MUP) and stores in it the produced assignment. Each mobile user that executed one or more tasks, by the end of these tasks, uploads the data to the external storage of the ISP and using a hash of them triggers the MUP smart contract that transfers the payment and the credits used for the calls of CSOPT and MUP. 

Table \ref{tab:interactions} lists and Figure \ref{fig:proto} depicts the interactions between the participating entities. Overall, four smart contracts are used. Two between the ISP and the CSPs and two between the ISP and the $MU$s. These contracts guarantee that \textit{(i)} the CSP will pay in order to get access to the collected data, \textit{(ii)} the mobile users will get paid if they do their tasks and will loose some credits if they will not, \textit{(iii)} the identity of the mobile users can not be revealed to the CSPs. 
Given that for each smart contract to be executed a transaction that has its address as a destination needs to be mined, it is worth mentioning that we assume that the mining time of a block in the blockchain is much shorter than the deadline of the crowdsensing request.

\subsection{Desirable Game Theoretic Properties of Auctions}\label{sec:desprop}

We need the mobile users to report their costs as well as the tasks they can do truthfully. If the payment scheme is not designed properly, as indicated in the following example, $MU$s can mis-report their bids to earn more money.

\textbf{Example:}~\textit{Challenges in the design of a truthful auction:}
Suppose there are 10 tasks and 3 interested $MU$s. 
$MU_1$ can do all these tasks at \$1 per task, $MU_2$ can do only task $T_{10}$ at \$1.5 and $MU_3$ can do all these tasks at \$2.5 per task.
If we decide to optimally select the set of $MU$s and pay them the first losing bid, all the tasks will be assigned to $MU_1$ who will be paid \$15 since the first losing bid is \$1.5 from $MU_2$ for $T_{10}$. However, $MU_1$ can misreport his bid to be \$1 per task but only for tasks $T_1$ to $T_9$. With this, he will obtain a payment of \$22.5 (2.5*9) since the first losing bid will be from $MU_3$ for tasks $T_1 - T_9$ and $MU_2$ will execute $T_{10}$ and earn \$2.5. The total cost in this case is \$25. Thus a careful design of the auction is necessary. 

If it is a best response for all the $MU$s to report their private information truthfully to an auction, we say the auction is \emph{incentive compatible}. We study auctions with respect to the following two notions of incentive compatibility.

	\textbf{(DSIC) Dominant Strategy Incentive Compatible:} An auction is called DSIC if reporting truthfully gives every $MU$ the highest utility regardless of the bids of the other $MU$s. 
    
    \textbf{(BIC) Bayesian Incentive Compatible:} An auction is called (BIC) if reporting truthfully gives an $MU$ highest expected utility when the other $MU$s are truthful, and the expectation is taken over bids of other $MU$s.  

Apart from incentive compatibility, we also need an auction to satisfy the \emph{individual rationality} property. 

	\textbf{(IR) Individually Rational:} An auction is called \emph{Individually Rational} (IR) if no $MU$ derives negative utility by participating in the auction.

Auctions could be designed with different goals. DSIC is a strong requirement that may be difficult to achieve. For that, it is a common approach in the design of auctions to enforce BIC and IR together with the desirable objective. The most popular objectives on the design of an auction are the auction to be \emph{Allocatively Efficient} or \emph{Cost Optimal}. An Allocatively efficient auction allocates the tasks to $MU$s having the least costs and achieves a socially good outcome while a cost optimal auction minimizes the cost incurred by the CSP. 

In crowd-sensing, it should be ensured that each task is completed with probability $\beta$ or higher. Let $r$ be a repeat factor, that is, each task is assigned to at least $r$ different users. The probability that the task is completed by at least one user is $1-(1-\alpha)^r \geq \beta$ or equivalently $r\geq \frac{\log (1 - \beta)}{\log (1 - \alpha)}$. We use $X_{ij}$ as an indication variable with $X_{ij}=1$ if $T_j$ is assigned to $MU_i$. Any auction on the examined setting needs to guarantee the following \emph{feasibility} conditions: \begin{eqnarray}
&\sum_i X_{ij} \geq \frac{\log (1 - \beta)}{\log (1 - \alpha)} \label{eq:repeat}\\
 &\{T_j\mid X_{ij}=1\} \subset \cT_i \; \forall i \label{eq:fesibility}
\end{eqnarray}

With this constraints, we define allocatively efficient (AE) and cost optimal (CO) auctions  as follows. 

\textbf{(AE) Allocatively Efficient Auction:} An auction that chooses assignments that minimize the total cost incurred by $MU$s for every reported cost. 

\textbf{Optimal Auction: } An auction that chooses assignments that minimize the total cost paid by the CSP.


DSIC, BIC, IR and AE are formally are formally defined in Section~\ref{sec:formalDefs}, while the optimal auction is discussed in the next Section. In order to design a BIC and IR auction, we also need to describe the conditions on the allocation rules and payments. 

\textbf{Truthfulness characterization: } Assuming that the cost per task is constant for all $MU$s. That is $\forall i \in \mathcal{N}$, $\mathbf{c}_i=(c_i,c_i,\ldots,c_i)$ 
and $c_i \in C_i = [\underline{c}_i, \bar{c}_i]$.
Let $n_i = \sum_j X_{ij}(\mathbf{b})$
The utility of a mobile user $i$ with bid $b_i$ is given as,
\begin{eqnarray*}
u_i({b}_i,{b}_{-i};\theta_i) &=& p_i - n_i c_{i} \\
U_i(b_i;\theta_i) &=& P_i(b_i) - c_{i} N_i(b_i) 
\end{eqnarray*}
where $N_i(b_i)$ is the expected number of tasks for $MU_i$ where the expectation is with respect to the bids of the other agents and $P_i(b_i)$ is the expected payment.\footnote{In general, the design of an optimal auction calls for designing the expected assignment and the expected payments for every user and every possible bid.}
We write, $P_i(b_i) = \rho_i(b_i) + \hat{c}_i N_i(b_i)$, where $\rho_i(b_i)$ is an additional incentive to report private information truthfully.
Thus, 
\begin{align}
U_i(b_i;\theta_i)
		&= \rho_{i}(b_i) -(c_i-\hat{c}_i)N_i(b_i) \label{eq:rho_utility}
\end{align}
Thus $\rho_i$ represents the offered utility when all the agents are truthful. With the above offered incentive, we have the following theorem. 
\begin{theorem}
\label{thm:bic_ir}
An auction is BIC and IR if and only if $\forall i \in \mathcal{N}$,
	\begin{enumerate}[leftmargin=0.4cm]
\itemsep0em 
		\item  	{$N_i(\hat{c}_i,\chT_i)$ is non-increasing in $\hat{c}_i  \forall \chT_i \subset \cT_i$}\label{thm:mon-cond2}.        
		\item {$\rho_{i}(b_i)$ is non-negative, and non-decreasing in $\hat{k}_i $  and $\forall\;\hat{c}_i\;\in\;[\underline{c}_i,\bar{c}_i]$}\label{thm:mon-cond1}		
		\item {$\rho_{i}(b_i) = \rho_{i}(\bar{c}_i,\hat{k}_i) + \int_{\hat{c}_i}^{\overline{c}_i}N_i(z,\hat{k}_i)dz $} \label{thm:utl-form}	
	\end{enumerate}
\end{theorem}
We refer to the above statements as conditions \ref{thm:mon-cond2}, \ref{thm:mon-cond1} and~\ref{thm:utl-form}.

\begin{pfof}
Though the key ideas in the proof are similar to~\cite{IYENGAR08,Myerson81}, note that our settings are quite different and we characterize the results in terms of $N_i$s and not $X_{ij}$s. We present the proof in Section~\ref{sec:proofs}.
\end{pfof}

\subsection{Ensuring the quality of Crowdsensing}

It is possible for some malicious mobile users to misreport the sensed data and affect their overall quality. This makes the building of a reputation system and a careful integration of reports in the final data necessary. There have been various approaches, such as in \cite{Radanovic16a,Resnick07}, proposed in the literature to limit the influence of low quality reporting. In particular, the Community Sensing Influence Limiter (CSIL) proposed by Radanovic and Faltings \cite{Radanovic16a} is the most suitable for our setting.
In CSIL, each $MU_i$ has a reputation score $\rho_i$ and data is added to the collected data with probability $\frac{\rho_i}{\rho_i+1}$. Thus, the influence of a malicious user on the aggregated data becomes limited. To build reputation scores, the ISP deploys certain trusted $MU$s across all cells. These $MU$s always perform the tasks assigned with honesty. Whenever, the ISP receives the data from trusted $MU$s, it updates the reputation score of each $MU$ who has reported data for that time slot. The reputation score update function captures how much the data supplied by $MU$ adds a value to the collected data. 

\subsection{Attack model and Defense}
In order to justify that our proposal preserves users' identities and location privacy, we design an attack from a CSP that wants to find them and we explain how it fails. In order for CSPs to identify the sensitive locations (home/work) of mobile users, they need to submit requests with short deadlines at times that they expect the participants to be at such locations. However, the ISP assigns a different temporary id to each participant every time. Even if the CSP submits the same request multiple times with a short deadline in a limited geographic area and even if it is always the same participant that completes the request, the ISP will preserve her privacy since she will be assigned a different randomly selected id every time. If the id is of the same length as the addresses in Ethereum (64 bytes), the range of the possible ids is $[1,2^{512}]$.

\section{Crowdsensing Optimal Auction}\label{sec:ISPandMUS}

For optimal request assignments, the true costs from the $MU$s are needed and hence we use \emph{mechanism design theory} to design auctions~\cite{nisan07chap9,Garg08a,Garg08b}. The goal can either be to minimize the cost incurred by the mobile users (AE auction) or to minimize the expected payment of the CSP (cost optimal auction). 
Note that,  the CSP's goal is not to care about game theoretic property, AE,  but to minimize its cost of such crowd sensing activity. Thus, we need to design a cost optimal auction for the CSP. 
In the examined setting, the mobile users bid for a certain set of desirable tasks and may get assigned its subset. In auction theory, this is called \emph{combinatorial auctions}. Designing optimal combinatorial auctions for general settings is an open problem. However, there have been different attempts for specific settings~\cite{Gujar09,Gujar13,Bhat15}. The key difference between ~\cite{Gujar09,Gujar13} and our settings is, in their paper a mobile user either assigned the set of tasks he is interested in or nothing where as in our settings, the mobile user may get subset of its desirable tasks.  In ~\cite{Bhat15}, the mobile user needs to submit capcity, that is how many tasks he can perform and the auction may assign any set of tasks not exceeding his capcity. In addition to combinatorial setting, we need to assign each task to multiple users to ensure high assurance on completion of tasks which is not addressed in the literature. Thus, the auction we design is categorized as an optimal multi-unit combinatorial auction. In general the characterization of an optimal combinatorial auction is an open problem. We leverage from the fact that although our setting is combinatorial, the tasks are homogeneous except from their locations. That is, a mobile user is indifferent to any constant size subset of tasks within its interested set of tasks. For example, a $MU$ who is interested in tasks $\mathcal{T}_1,\mathcal{T}_2,\mathcal{T}_3,\mathcal{T}_4$, incurs the same cost if it is assigned $\mathcal{T}_1,\mathcal{T}_2$ or $\mathcal{T}_3,\mathcal{T}_4$ or any two of these fours tasks.

We start designing an optimal auction with game theoretic properties BIC and IR. With our BIC and IR characterization result, we provide sufficient conditions for an auction to be an optimal auction in our context. 
Next, we study the concept of \textbf{Regularity} and prove that the optimal auction we have designed is also AE under regularity.  
Then we design a payment rule which along with AE allocation rule qualifies to be an optimal auction. The proposed payment rule offers difference between the cost of AE auction with their presence and absence as incentives to report their costs truthfully. That is if the cost of a $MU$ is \$5 and the AE cost increases in his absence by \$2, it is paid \$7. We design an efficient allocation rule to determine allocation rule satisfying AE property (Algorithm 1, subroutine ALLOC-RULE). We call the proposed auction as \emph{CSOPT}. 
Note that, though we set the goal to design an optimal auction with BIC and IR as constraints, CSOPT along with cost optimality also satisfies AE and DSIC. 

\subsection{\MECH: Cost Optimal Mobile Crowdsensing Auction}

An auction is called optimal, for CSP, if it minimizes the total expected payment to the $MU$s, is BIC and IR and is feasible~\cite{Myerson81}. That~is: 
\begin{eqnarray}
\mbox{minimize} &\mathbb{E}_{\mathbf{b}} \sum_{i\in \mathcal{N}}  p_i(\mathbf{b}) &\nonumber\\
\mbox{subject to: BIC} & \qquad U_i(c_i,\cT_i;\theta_i)\geq U_i(b_i;\theta_i) \forall c_i,\forall \cT_i  \nonumber\\
\mbox{IR} &U_i({c}_i,\cT_i;\theta_i)\geq 0\nonumber\\
\mbox{FEASIBILITY} & \sum_i X_{ij} \geq \frac{\log (1 - \beta)}{\log (1 - \alpha)} \nonumber\\
\mbox{FEASIBILITY} & \{ T_j\mid X_{ij}= 1 \} \subset \cT_i \; \forall i  \nonumber
\end{eqnarray}
Let $F_i(c_i|k_i)$ and $f_i(c_i|k_i)$ denote respectively the cumulative distribution and probability density function of cost ($c_i$) of  $MU_i$ given the number of tasks it can perform. 
\begin{theorem}
\label{thm:offline_payment}
Suppose the allocation rule minimizes
\begin{align}
\label{eq:opt}
&\sum_{i=1}^{n} \int_{\underline{c}_1}^{\bar{c}_1}\ldots \int_{\underline{c}_n}^{\bar{c}_n}  \bigg(c_i + \frac{F_i(c_i|k_i)}{f_i(c_i|k_i)} \bigg)  \nonumber n_i(c_i,k_i,c_{-i},k_{-i}) \\ & f_1(c_1,k_1) \ldots f_n(c_n,k_n) \,dc_1\ldots dc_n dk_1dk_2 \ldots dk_n
\end{align} $\forall k_i$
\normalsize
\noindent subject to conditions \ref{thm:mon-cond2} and \ref{thm:mon-cond1} of 
Theorem \ref{thm:bic_ir}, Equation ~(\ref{eq:repeat}) and Equation~(\ref{eq:fesibility}). Also, suppose the payment is given by
\begin{align}
P_i(c_i,k_i) = c_iN_i(c_i,k_i) + \int_{c_i}^{\overline{c}_i}
	N_i(z,k_i)dz \label{eqn:opt_payment}
\end{align}
\noindent then such a payment scheme and allocation scheme constitute an optimal auction satisfying BIC and IR.
\end{theorem}
\begin{pfof}
The proof is given in Section~\ref{sec:proofs}.
\end{pfof}

\textbf{(Regularity):} We define the virtual cost function as
\begin{align*}
H_i(c_i,k_i) := c_i + \frac{F_i(c_i|k_i)}{f_i(c_i|k_i)}, \forall MU_i \in \mathcal{N}
\end{align*}
We say that a type distribution is regular if $\forall i$, $H_i$ is non-decreasing in $c_i$ and non-increasing in $k_i$. Analogous to the literature on optimal auctions~\cite{IYENGAR08,Myerson81}, we assume regularity on our distribution type. We assume the type distributions satisfy regularity and all the $MU$ types are independently and identically distributed (i.i.d.) over $ [\underline{c_l},\overline{c_u}] \times  [\underline{k_l},\overline{c_u}] $. We make a further assumption that the costs for all $MU$s are identically distributed. With these assumptions, we present the pseudocode of \MECH\ in Algorithm~\ref{alg:mech}. 

\begin{algorithm}[t]
\caption{\sc  \MECH }
\label{alg:mech}
\KwIn{ $\mathcal{N},\mathcal{T},\mathbf{\hat{c}},(\chT_i)_{i\in\mathcal{N}}, r$ }
\KwOut{Allocations $\mathcal{A}=(\mathcal{AT}_1,\mathcal{AT}_2,\ldots,\mathcal{AT}_n)$  and Payments $\mathcal{P} = \{p_1,p_2,\ldots,p_n\}$.}
\textbf{Allocations:} \\
$\cT \leftarrow r \cT$ // Make $r$ copies of each task in $\cT$\\
$\mathcal{A}$ = ALLOC-RULE($\mathcal{N},\cT,\mathbf{\hat{c}},(\mathcal{T}_i)_{i\in\mathcal{N}}$)\\

$[p_1,p_2,\ldots,p_n]$ = PAYMENT-RULE($\mathcal{N},\cT,\mathbf{\hat{c}},(\mathcal{T}_i)_{i\in\mathcal{N}}, \mathcal{A}$)\\
\label{alg:tstwo}
\setcounter{AlgoLine}{0}
\hrule
\vspace{0.1cm}
Subroutine: ALLOC-RULE($\mathring{N},\mathring{\cT},\mathring{\mathbf{c}},(\mathring{\cT}_i)_{i\in \mathring{N}}$) \nonumber \;
\hrule
\vspace{0.1cm}
\KwIn {
 $\langle \mathring{N},\mathring{\cT},\mathring{\mathbf{c}},(\mathring{\cT}_i)_{i\in \mathring{N}} \rangle$
}
\KwOut {Vector $\mathcal{A}= (\mathcal{AT}_1,\mathcal{AT}_2,\ldots,\mathcal{AT}_n)$ tasks assigned to each $MU$.}
\While{$\mathring{\cT} \neq \emptyset$}{
Sort $MU$s based on cost per task for tasks in~$\mathring{\cT}$\\
Add the most economic $MU$, say $MU_i$ to the selected $MU$s\\
 $\mathcal{A}\mathcal{T}_i = \mathring{\cT}\cap\mathring{\cT}_i $\\
$\mathring{\cT} \leftarrow \mathring{\cT} - \mathcal{A}\mathcal{T}_i$\\
}
$\mathcal{A}=(\mathcal{A}\mathcal{T}_1,\mathcal{A}\mathcal{T}_2,\ldots,\mathcal{A}\mathcal{T}_n)$
\hrule
\vspace{0.1cm}
Subroutine: PAYMENT-RULE($\mathcal{N},\cT,\mathbf{\hat{c}},(\mathcal{T}_i)_{i\in\mathcal{N}}, \mathcal{A}$) \nonumber \;
\hrule
\vspace{0.1cm}
\KwIn {
 $\langle \mathcal{N},\cT,\mathbf{\hat{c}},(\mathcal{T}_i)_{i\in\mathcal{N}}, \mathcal{A} \rangle$ 
}
\KwOut {Vector $\mathcal{P}$ of payments of each agent.}
	fcost = COST($\mathcal{A}$)\\
    // COST finds out the cost of allocation $\mathcal{A}$ \\
\For{$j\in \mathcal{N}$}
{
	$\mathring{A}_j$=ALLOC-RULE($\mathcal{N}\setminus\{j\},\cT,\mathbf{c}_{-j},(\chT_i)_{i\in \mathcal{N}\setminus\{j\}} $)\\
	scost=COST($\mathring{A}_j$)\\
	$p_j$= $n_j\times c_j$ + scost-fcost\\
}
\end{algorithm}

\begin{observation}
Under the assumption of regularity and i.i.d. $MU$s, an allocatively efficient auction  is an optimal solution to Equation~(\ref{eq:opt}) and maximizes Equation~(\ref{eq:opt}) for each $\mathbf{b}$.
\end{observation}
\begin{observation}
Under the assumption of regularity and i.i.d. $MU$s, for a fixed $b_{-i}$, the following payment satisfies Equation~(\ref{eqn:opt_payment}).
\begin{align}
\label{eqn:pay_mech}
p_i(c_i,k_i,b_{-i}) = c_i n_i(c_i,k_i,b_{-i}) + \int_{c_i}^{\overline{c}} n_i(z,k_i,b_{-i})dz
\end{align}
\end{observation}
Since we are using an AE allocation, the payment (\ref{eqn:pay_mech}) can be written as:
\begin{equation*}
    p_i() = c_i n_i(c_i,k_i,b_{-i}) + V_{-i}^* -  V^* 
\end{equation*}
where $V^*$ is the cost of AE allocation and $V_{-i}^*$ is the cost of AE allocation if $MU_i$ is not 
in the system. Observe that, keeping $b_{-i}$ fixed, whenever $MU_i$ increases its cost, either $n_i()$ remains the same or drops by some integer until it goes to zero. Let us assume $c_i < c_{i1} < \ldots c_{il} <\overline{c}$ are the costs at which $n_i$ drops. Since we assume there is enough competition, eventually it should drop to zero that is $n_i(c_{il},k_i.b_{-i})=0$. Precisely $c_{i1},c_{i2},\ldots.c_{il}$ are the costs which get added into an AE allocation when $MU_i$ is not there in the system.

With all the above discussion, we propose our mechanism \MECH\ as given in Algorithm \ref{alg:mech}. COST($\mathcal{A}$) returns the total cost of allocating tasks as described in $\mathcal{A}$.
Hence $scost$ captures the total cost incurred by $MU_j$ in optimal allocation.
\begin{lemma}
\label{lemma:ae}
\MECH\ is an AE auction for the CSP.
\end{lemma}
\begin{pfof}
By construction, it satisfies FEASIBILITY conditions. We need to show that it minimizes the total allocation cost. Let $\mathcal{A}_e$ be an AE allocation given bids as $\mathbf{b}$. Let $MU_1,MU_2,\ldots$ be the order in which \MECH\ allocates the tasks to $MU$s. Let $MU_i$ be the first $MU$ whose allocation in \MECH\ differs from that of in $\mathcal{A}_e$. Thus at least one of its tasks is assigned to $MU_j$ $j>i$. However, $c_i \leq  c_j$. Thus not awarding all $n_i()$ tasks to $MU_i$ which are allocated by \MECH\ the cost is not going to improve. Using induction, it follows that no allocation $\mathcal{A}_e$ can improve on cost of allocation over \MECH. 
\end{pfof}

\begin{theorem}
\MECH\ is an optimal auction for the CSP.
\end{theorem}
\begin{pfof}
It follows from Observations 1,2, Lemma \ref{lemma:ae} and Theorem \ref{thm:offline_payment}.
\end{pfof}


\begin{figure*}[t]
\centering
\begin{subfigure}{0.63\columnwidth}
	\centering
	\includegraphics[width=\columnwidth0]{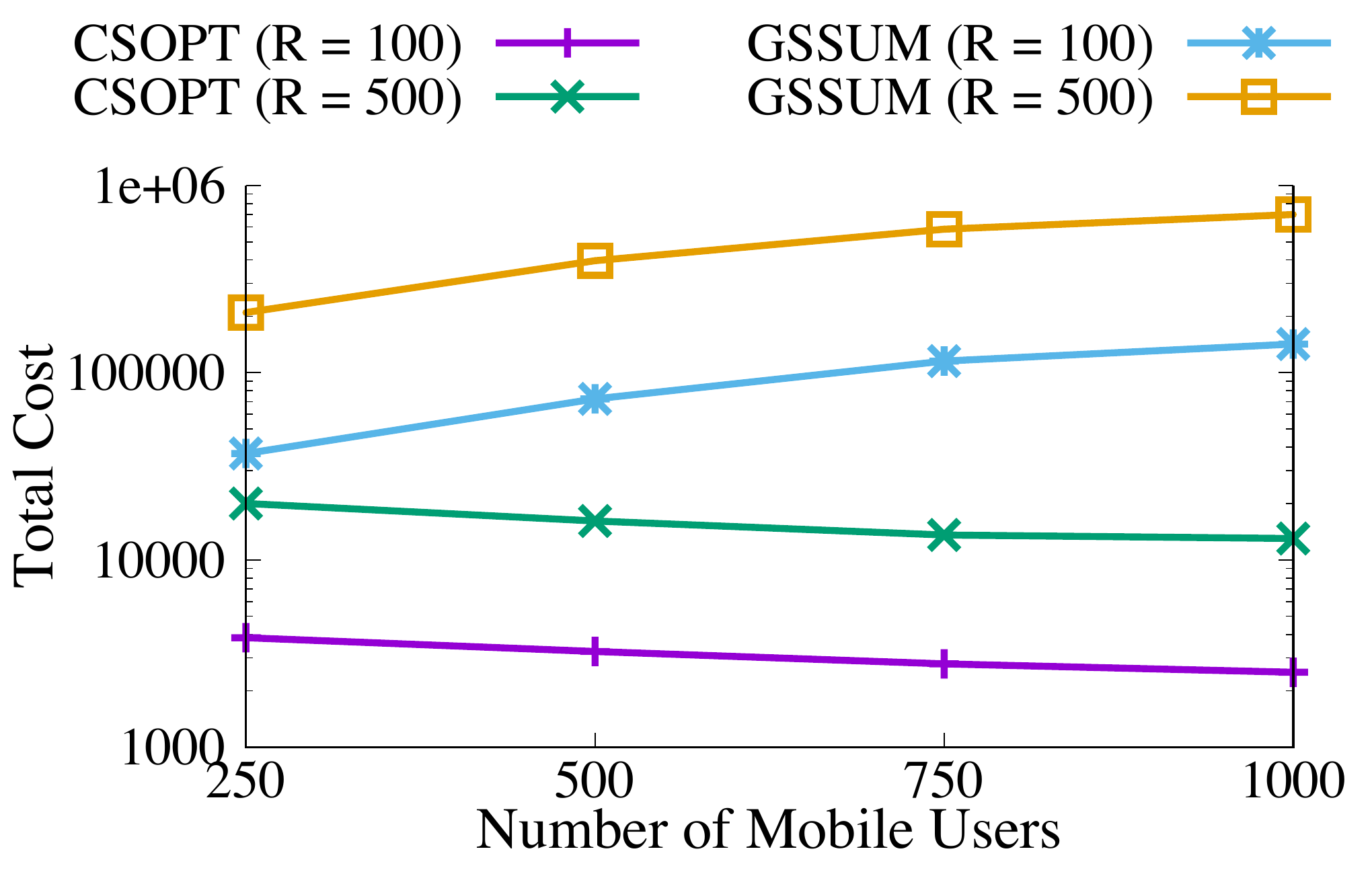}
    \caption{Auction cost as function of the number of the mobile users.}
    \label{fig:mu}
\end{subfigure}
\hfill
\begin{subfigure}{0.63\columnwidth}
	\centering
	\includegraphics[width=\columnwidth]{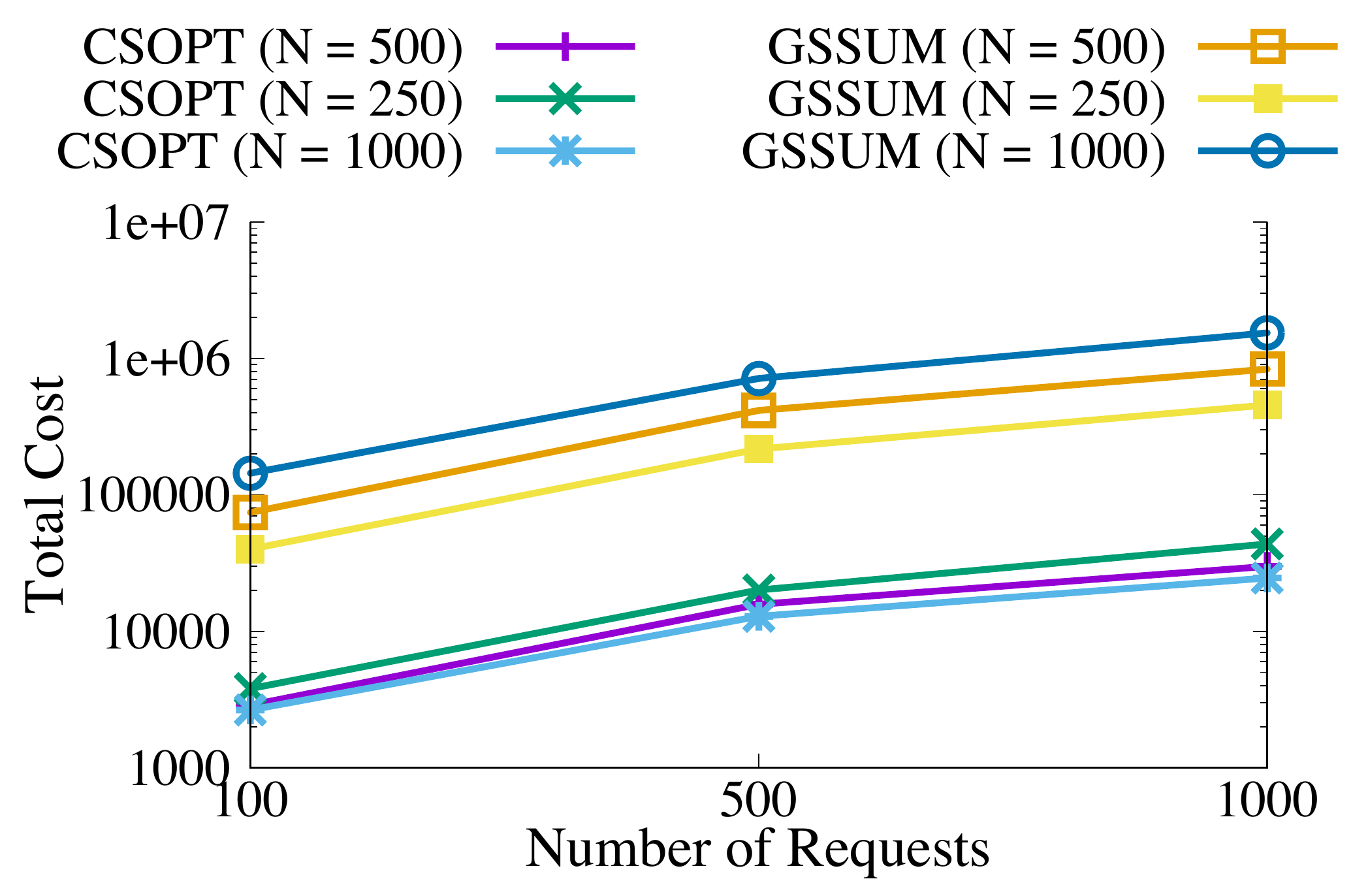}
    \caption{Auction cost as a function of the number of the requests.}
    \label{fig:reqs}
\end{subfigure}
\hfill
\begin{subfigure}{0.63\columnwidth}
	\centering
	\includegraphics[width=\columnwidth]{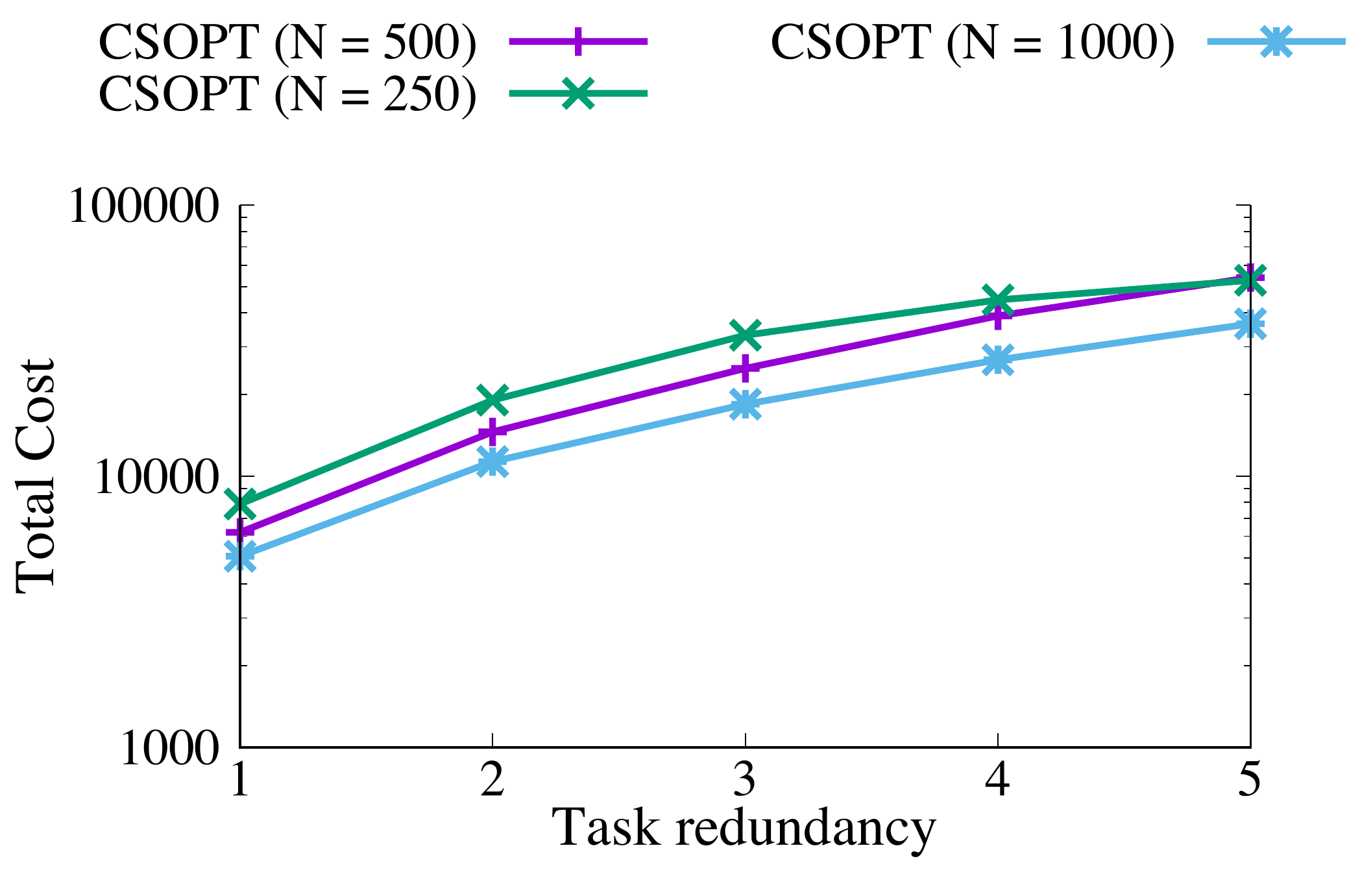}
    \caption{Auction cost as a function of the task redundancy.}
    \label{fig:r}
\end{subfigure}
\caption{Performance evaluation of CSOPT and comparison with GSSUM algorithm that was proposed by the authors of~\cite{Karaliopoulos15}.}
\label{fig:auction}
\end{figure*}

\begin{figure*}[t]
\centering
\begin{subfigure}{0.99\columnwidth}
	\centering
	\includegraphics[height=0.76\columnwidth,angle=-90]{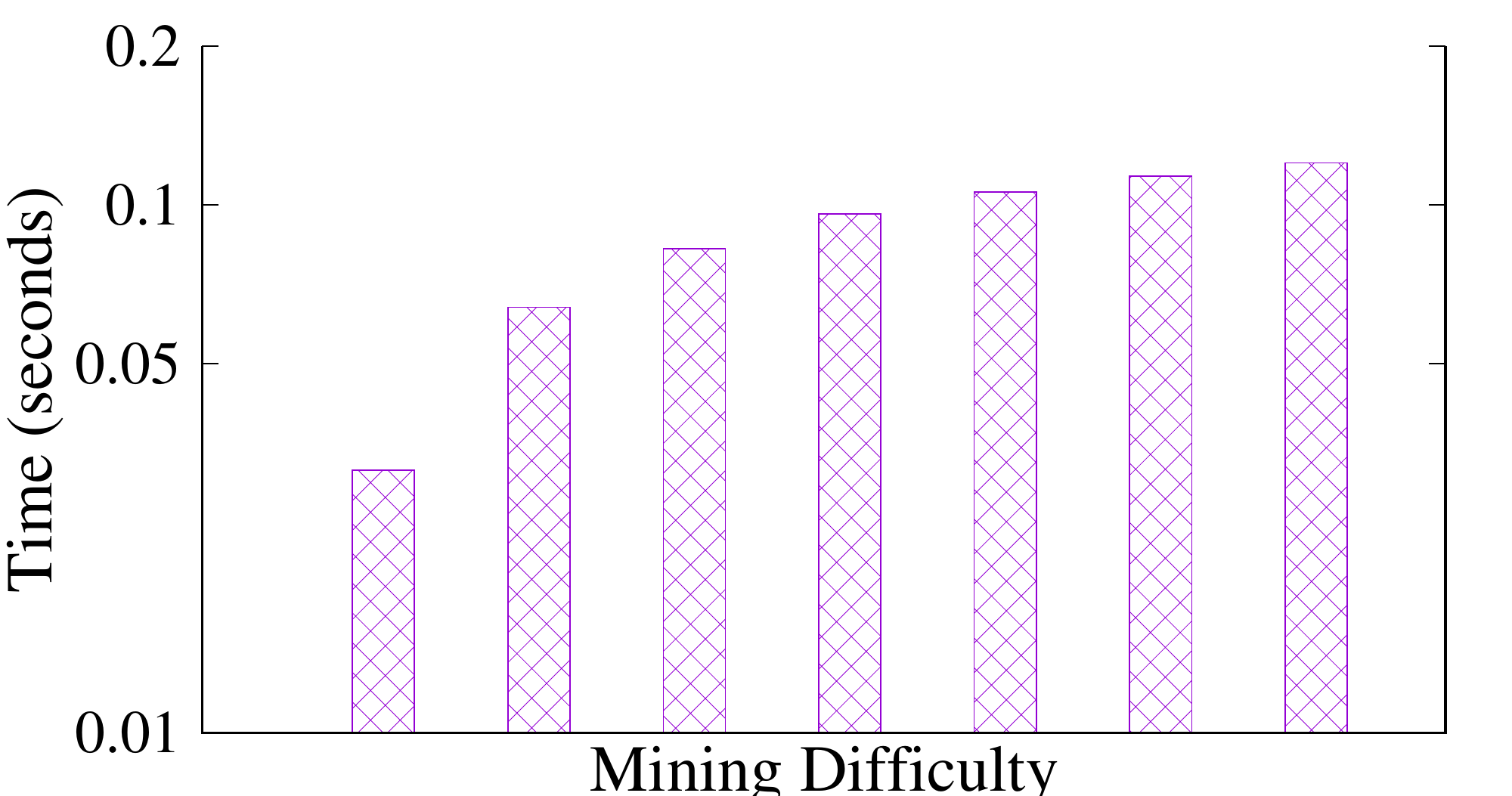}
    \caption{The mining time of a new block for a given mining difficulty.}
    \label{fig:mining}
\end{subfigure}
\begin{subfigure}{0.99\columnwidth}
	\centering
	\includegraphics[height=0.76\columnwidth,angle=-90]{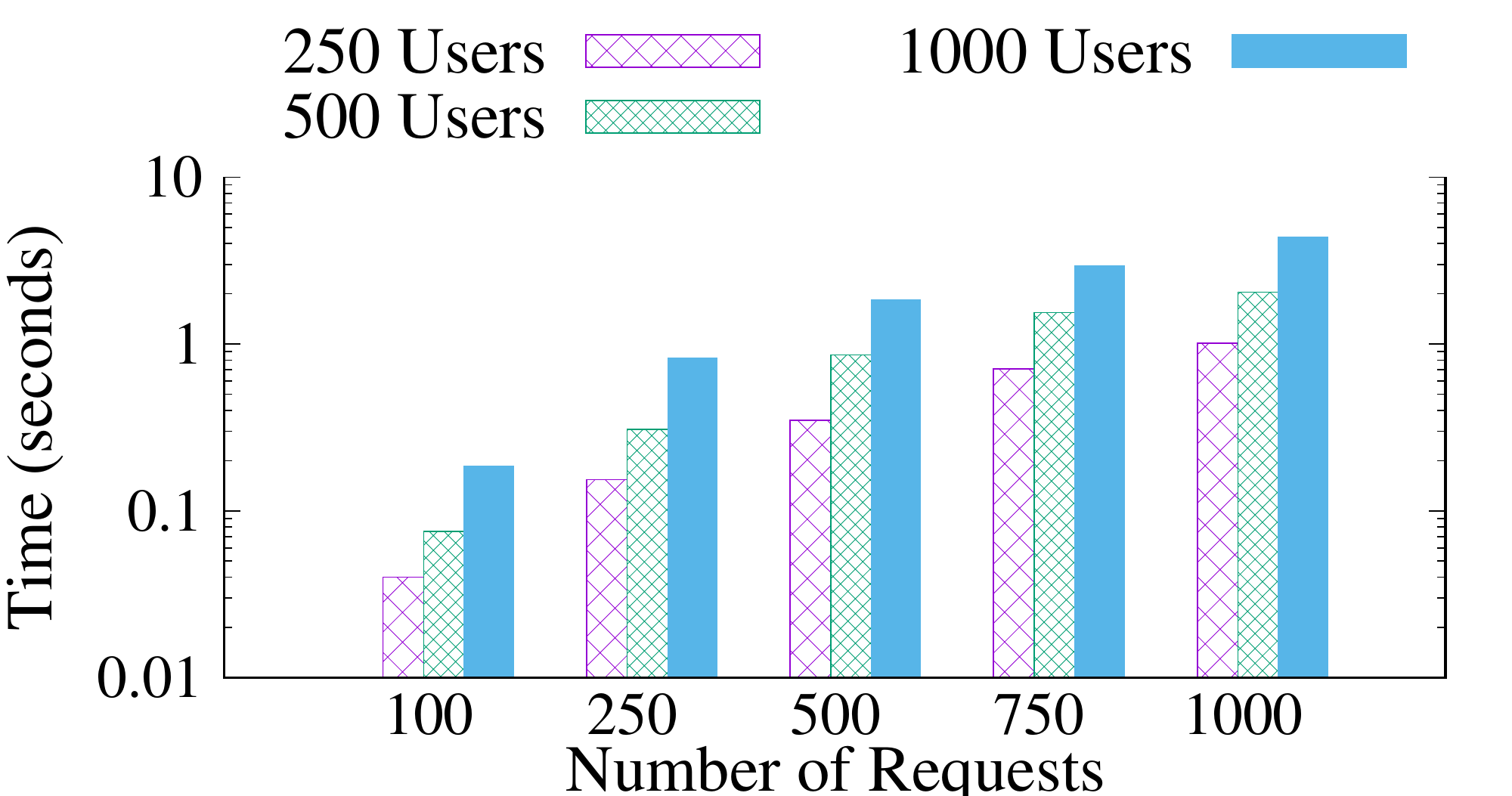}
    \caption{The time needed for the execution of CSOPT for different number of users and requests.}
    \label{fig:smartContract}
\end{subfigure}
\caption{The two components that affect execution time. Mining difficulty determines the time needed for the execution of a smart contract. CSOPT cannot be executed instantaneously and for that we measure its time requirements separately.}
\label{fig:archTimes}
\end{figure*}

\section{Performance Evaluation}\label{sec:evaluation}

We conduct two set of experiments to depict the quality of our proposal. First, we compare the cost of the proposed auction with another state-of the-art algorithm. Next, we examine the time needs of the architecture to process crowdsensing requests. Figure~\ref{fig:auction} shows the results from the first set of experiments and Figure~\ref{fig:archTimes} of the second.

\textbf{CSOPT:} Since we have proved mathematically that \MECH\ is allocatively efficient we only need to compare the cost of the allocations it produces with other state-of-the-art algorithms. For that we selected~\cite{Karaliopoulos15} because it adopts to our scenario and it is also fast in terms of time since it operates in a greedy manner. The authors named their algorithm greedy heuristic for selection under stochastic user mobility, but here we refer to it as GSSUM for short. We consider an area that is composed by a 100 by 100 grid and we randomly place mobile users on this grid. Then we generate crowdsensing requests with deadlines and we assume that each user bids for a task only if it is within a certain distance. Each user has a cost per allocated task in a range between 50 and 100. Figure \ref{fig:mu} shows, in logarithmic scale, that the total cost (payments to the $MU$s) of CSOPT is decreasing as the number of $MU$s is increasing. This result is expected because the increased competition between $MU$s decreases the cost per task. On the other hand, the GSSUM algorithm operates in the opposite way because whenever it selects a mobile user to assign a task to, it assigns all the tasks on which she has a bid. 

Next, we compare the two algorithms in terms of the number of requests. Figure~\ref{fig:reqs} shows that CSOPT is at least one order of magnitude less costly while the cost in both algorithms is increasing at the same rate. Next, in Figure \ref{fig:r} we show how the costs increase when the number of requests reaches 200 but there is a repeat factor $r=[1,2,3,4,5]$ that ensures that the requests will be satisfied, as explained in Section~\ref{sec:desprop}. This case differs from the previous one because the requests are less disseminated throughout the whole area and the average number of the participants that can handle a request is much smaller. $r$ in Figures~\ref{fig:mu}~and~\ref{fig:reqs} is 1. 


\textbf{Architecture:} We install Ethereum in a Desktop with Intel Core i7-7700 CPU @ 3.60 GHz and 16 GB of RAM. We then measure the time required for a block to be mined for different values of mining difficulty and the time requirements of CSOPT. Figure \ref{fig:archTimes} depicts these measurements. In order to produce Figure \ref{fig:mining} we set the mining difficulty on the genesis file of Ethereum and wait for 100 blocks to be mined. Small values of mining difficulty can produce a new block every few millisecond but this will produce the generation of many empty blocks that are a waste of storage.
For the time measurements of CSOPT for different numbers of $MU$s and crowdsensing requests, we implement the algorithm and measured its performance on the same desktop. Figure \ref{fig:smartContract} shows that the time CSOPT needs to determine the task assignment and the payments increases with the number of requests and $MU$s. However, it does not require more than 10 seconds in the case of 1000 $MU$s and 1000 tasks. We denote the time requirements of CSOPT with $t_{CSOPT}$ and the mining time of a block by $t_{B}$.

These experiments are important since all the blockchain-based interactions as described in Section \ref{sec:blockchain} will have this delay. In total, any request from a CSP needs: 1 block to be mined in order to register the request (RR) while in parallel the ISP contacts the mobile users and announces the tasks ($t_{ann}$). If the announcement time takes more time than the mining of RR, the mining time of this block can be ignored. Next, the users bid for a predefined time period ($t_{bidding}$) and after that the CSOPT is triggered, whose termination triggers MUP. If the crowdsensing task duration $t_{task}$ is longer than $t_{B}$, the mining of the block that is caused by the MUP is not counted in the total delay. By the end of the task execution, the users submit the collected data and DA is triggered after $t_B$ notifies the CSP that the data have been collected. The total delay between the submission of the request and the access to the collected data is: 
\begin{equation*}
\max(t_{B},t_{ann}) + t_{bidding} + t_{CSOPT}+\max(t_{B},t_{task}) + t_{B}.
\end{equation*}
From this set of experiments we can conclude that if the duration of the crowdsensing tasks is in the order of tens of seconds the time overhead of using a blockchain instead of a centralised server is negligible while the benefits in terms of preserving the users privacy are high.

\section{Conclusion}\label{sec:conclusion}
In conclusion, we proposed a novel architecture for event-based spatial crowdsensing tasks that is deployed by an ISP and is based on blockchain technology. The proposed architecture employs smart contracts \textit{(i)} that allow crowdsensing service providers to submit their requests, \textit{(ii)} to run a cost-optimal auction for the determination of the most suitable mobile users that are interested in executing the crowdsensing tasks, \textit{(iii)} to deal with the payments for the mobile users and \textit{(iv)} to give access to the crowdsensing provider. The proposed architecture preserves the privacy of the mobile users in the sense that the crowdsensing provider cannot know their identity and can not derive their sensitive information such as the location of their home/work. Moreover, we have shown that the employed incentive compatible cost optimal auction that determines the selection of the mobile users that will handle each crowdsensing task, outperforms state of the art proposals when adopted to the examined setting by one order of magnitude for high numbers of mobile users and tasks.

\section{Acknowledgements}
This research has been supported, in part, by projects 26211515 and 16214817 from the Research Grants Council of Hong Kong.

\bibliographystyle{IEEEtran}
\bibliography{PPASC}



\section{Formal Definitions}\label{sec:formalDefs}

\noindent\begin{definition}{\bf{(Dominant Strategy Incentive Compatible):}}

\noindent An auction is called \emph{Dominant Strategy Incentive Compatible} (DSIC) if reporting truthfully gives every $MU$ the highest utility regardless of the bids of the other $MU$s. 

Formally, $\forall i \in N, \forall \mathbf{c}_i, \hc_i \in\mathbf{C}_i \forall \mathcal{\hat{T}}_i \subset \mathcal{T}_i$, $\forall {b}_{-i}$,
\begin{align*}
 u_i(\mathbf{c}_i,\cT_i,\mathbf{b}_{-i};\theta_i) \geq u_i(\hc_i,\chT_i,\mathbf{b}_{-i};\theta_i).
 \end{align*}

\end{definition}

\noindent\begin{definition}{\bf{(Bayesian Incentive Compatible):}}

\noindent An auction is called \emph{ Bayesian Incentive Compatible} (BIC) if reporting truthfully gives an $MU$ highest expected utility when the other $MU$s are truthful, and the expectation is taken over bids of other $MU$s. 

Formally, $\forall i \in \mathcal{N}, \forall \mathbf{\hat{c}_i}, \mathbf{c_i}$, 
\begin{align}
\nonumber U_i(\mathbf{c_i},\mathcal{T}_i;\theta_i) \geq U_i(\mathbf{\hat{c_i}},\mathcal{\hat{T}}_i;\theta_i),
 \end{align}
 where, $U_i(\mathbf{b}_i;\theta_i) = \mathbb{E}_{b_{-i}}[u_i(\mathbf{b}_i,\mathbf{b}_{-i};\theta_i)]$.
\end{definition}

\begin{definition}{\bf{(Individually Rational):}}

\noindent  An auction is called \emph{Individually Rational} (IR) if no $MU$ derives negative utility by participating in the auction. 

Formally, $\forall i \in N, \forall \mathbf{c}_i \in \mathbf{C}_i, \cT_i \subset \cT$,
\begin{align*}
u_i(\mathbf{c_i},\mathcal{T}_i, \mathbf{b}_{-i};\mathbf{c_i},\mathcal{T}_i) \geq 0
\end{align*}
\end{definition}

\begin{definition}{\bf{(Allocatively Efficient (AE) Auction):}}\label{def:ae}

\noindent If an auction chooses assignments that minimize the total cost incurred by $MU$s for every reported cost, we call it an \emph{allocatively efficient (AE)} auction. 

That is, $\forall \mathbf{c}$ the auction assigns tasks such that: 
\begin{eqnarray}
\underset{\mathbb{X}}{\text{minimize}} &\sum_{i\in \mathcal{N}} \sum_{j=1}^{j=k}c_{ij}X_{ij} \label{eq:ae}\\
\text{subject to}  &\sum_i X_{ij} \geq \frac{\log (1 - \beta)}{\log (1 - \alpha)} \label{eq:repeat2}\\
 &\{T_j\mid X_{ij}=1\} \subset \cT_i \; \forall i \label{eq:fesibility2}
\end{eqnarray}
and each task is assigned to at least $r$ different mobile users. 
\end{definition}

\section{Proofs}\label{sec:proofs}
\subsection{Proof of Theorem \ref{thm:bic_ir}}
\begin{pfof}
\noindent To prove the necessity part of the theorem, we first observe due to BIC we have, 
\begin{align*}
&U_i(\hat{c}_i,\hat{k}_i;c_i,k_i) \leq  U_i(c_i,k_i;c_i,k_i) \qquad\forall(\hat{c}_i,\hat{k}_i) \mbox{ and }(c_i,k_i)\\
&\implies U_i(\hat{c}_i,k_i;c_i,k_i)\leq U_i(c_i,k_i;c_i,k_i)
\end{align*}
Without loss of generality, we assume $\hat{c}_i>c_i$ 
Rearrangement of these terms yields,
\begin{align*}
U_i(\hat{c}_i,k_i;c_i,k_i) = U_i(\hat{c}_i,k_i;\hat{c}_i,k_i) 
+ (\hat{c}_i-c_i)N_i(\hat{c}_i,k_i),
\end{align*}
which implies that,
\begin{align*}
\frac{U_i(\hat{c}_i,k_i;\hat{c}_i,k_i)-U_i(c_i,k_i;c_i,k_i)}{\hat{c}_i-c_i}
 \leq -N_i(\hat{c}_i,k_i).
\end{align*}

Similarly using $U_i(c_i,k_i;\hat{c}_i,k_i) \leq U_i(\hat{c}_i,k_i;\hat{c}_i,k_i)$,
\begin{align}
-N_i(c_i,k_i) &\leq\frac{U_i(\hat{c}_i,k_i;\hat{c}_i,k_i)-U_i(c_i,k_i;c_i,k_i)}{\hat{c}_i-c_i}\nonumber \\
&\leq-N_i(\hat{c}_i,k_i).\label{eq:mono1}
\end{align}
\noindent Taking limit $\hat{c}_i\rightarrow c_i,$ we get,
\begin{align}
\frac{\partial U_i(c_i,k_i;c_i,k_i)}{\partial{c}_i} = -N_i(c_i,k_i). 
\label{eq:pde}
\end{align}
Equation (\ref{eq:mono1}) implies, $N_i(c_i,k_i)$ is 
non-increasing in $c_i$.  This proves condition \ref{thm:mon-cond2} of the 
theorem in the forward direction. When the worker bids truthfully, 
from Equation (\ref{eq:rho_utility}),
\begin{align}\label{eq:rho1}
\rho_{i}(c_i,k_i)=U_i(c_i,k_i;c_i,k_i).
\end{align}
For BIC, Equation (\ref{eq:pde}) should be true. So,
\begin{align}
\rho_{i}(c_i,k_i)=\rho_{i}(\bar{c}_i,k_i)+\int_{c_i}^{\bar{c}_i}N_i(z,k_i)dz\label{eq:rho2}
\end{align}
This proves condition \ref{thm:utl-form} of the theorem. BIC also requires,
\begin{align*}
k_i \in \mbox{argmax}_{\hat{k}_i} 
U_i(c_i,\hat{k}_i;c_i,k_i)
\;\forall\; c_i\;\in\;[\underline{c}_i,\bar{c}_i]
\end{align*}

\noindent This implies, $\forall c_i,\;\rho_{i}(c_i,k_i)$ should be 
non-decreasing in $k_i$. The IR conditions (Equation(\ref{eq:rho1})) imply \begin{equation*}
\rho_{i}(c_i,k_i)\geq 0.
\end{equation*}
This proves condition \ref{thm:mon-cond1} of the theorem. Thus, these three 
conditions are necessary for BIC and IR properties.
We now prove the sufficiency. Consider
\begin{align*}
U_i(c_i,k_i;c_i,k_i)=\rho_i(c_i,k_i) \geq 0.
\end{align*}
So the IR property is satisfied. Without loss of generality, we assume $\hat{c}_i>c_i.$ The proof is similar for the case $\hat{c}_i<c_i.$
\begin{align*}
&U_i(b_i;c_{i},k_i) \\
&=\rho_{i}(\hat{c}_i,\hat{k}_i)+(\hat{c}_i-c_i)N_i(\hat{c}_i,\hat{k}_i)\tag*{(By Defn)}\nonumber \\
&= \rho_{i}(\bar{c}_i,\hat{k}_i) 
		+ \int_{\hat{c}_i}^{\bar{c}_i}N_i(z,\hat{k}_i)dz 
		+ (\hat{c}_i-c_i)N_i(\hat{c}_i,\hat{k}_i) \tag*{(By hypothesis)} \nonumber \\
&= \rho_{i}(\bar{c}_i,\hat{k}_i) 
		+ \int_{c_i}^{\bar{c}_i}N_i(z,\hat{k}_i)dz \\
& \qquad \qquad - \int_{c_i}^{\hat{c}_i}N_i(z,\hat{k}_i)dz 
		+ (\hat{c}_i-c_i)N_i(\hat{c}_i,\hat{k}_i)\nonumber \\
&\leq \rho_{i}(c_i,\hat{k}_i)
		\tag*{($N_i$ is non-increasing in $c_i$)}
	\nonumber \\
&\leq \rho_{i}(c_i,k_i) \tag*{( as $\rho_{i}$ is non-decreasing in $k_i$)} \nonumber \\
&= U_i(c_{i},k_i;c_i,k_i) 		 \nonumber
\end{align*}
\end{pfof}

\subsection{Proof of the Theorem \ref{thm:offline_payment}}

\begin{pfof}
The auctioneer's objective is to maximize her expected utility subject to conditions BIC, IR, and Feasibility. Her objective function is:
\begin{align}
&\sum_{i=1}^{n}\int_{\underline{c}_1}^{\bar{c}_1}\ldots \int_{\underline{c}_n}^{\bar{c}_n}\int_{\underline{k}_1}^{\bar{k}_1} \ldots\int_{\underline{k}_n}^{\bar{k}_n} \big[  -p_i(b)\big] \nonumber \\
& f_1(c_1,k_1)\ldots f_n(c_n,k_n) dc_1\ldots dc_n \, dk_1 \ldots dk_n  \nonumber \\
&=\sum_{i=1}^{n}\int_{\underline{c}_1}^{\bar{c}_1}\ldots \int_{\underline{c}_n}^{\bar{c}_n}\int_{\underline{k}_1}^{\bar{k}_1} \ldots\int_{\underline{k}_n}^{\bar{k}_n} \big[ ( -c_i + c_i)n_i(c_i,k_i,c_{-i},k_{-i})-p_i(b)\big]  \nonumber \\
&\qquad f_1(c_1,k_1)\ldots f_n(c_n,k_n) dc_1\ldots dc_n \, dk_1 \ldots dk_n  \nonumber \\
&=\sum_{i=1}^{n}\int_{\underline{c}_1}^{\bar{c}_1}\ldots \int_{\underline{c}_n}^{\bar{c}_n}\int_{\underline{k}_1}^{\bar{k}_1} \ldots\int_{\underline{k}_n}^{\bar{k}_n} \big( -c_i n_i(c_i,k_i,c_{-i},k_{-i}) \big)\nonumber \\
 & f_1(c_1,k_1)\ldots f_n(c_n,k_n) dc_1\ldots dc_n \, dk_1 \ldots dk_n  \nonumber \\
&+\sum_{i=1}^{n} \int_{\underline{c}_1}^{\bar{c}_1}\ldots \int_{\underline{c}_n}^{\bar{c}_n}\int_{\underline{k}_1}^{\bar{k}_1} \ldots\int_{\underline{k}_n}^{\bar{k}_n} \Bigg(  c_i n_i(c_i,k_i,c_{-i},k_{-i}) -p_i(b) \Bigg) \nonumber \\
&\qquad  f_1(c_1,k_1) \ldots f_n(c_n,k_n) \,dc_1\ldots dc_n \, dk_1 \ldots dk_n  \label{opt_stmtint}
\end{align}
The first term of Equation~(\ref{opt_stmtint}) is already same as first term in desired form of objective function of auctioneer given in Equation~(\ref{eq:opt}). We now use conditions (\ref{thm:mon-cond2}) and (\ref{thm:utl-form}) of Theorem \ref{thm:bic_ir} to arrive at the result.
{\allowdisplaybreaks
\begin{align}
&\int_{\underline{c}_1}^{\bar{c}_1}\ldots \int_{\underline{c}_n}^{\bar{c}_n}\int_{\underline{k}_1}^{\bar{k}_1} \ldots\int_{\underline{k}_n}^{\bar{k}_n} \big( c_i n_i(.)-p_i(b) \big) \nonumber \\ &f_1(c_1,k_1)\ldots f_n(c_n,k_n) dc_1\ldots dc_n \, dk_1 \ldots dk_n  \nonumber \\
&= - \int_{\underline{k}_i}^{\bar{k}_i}\int_{\underline{c}_i}^{\bar{c}_i} \rho_i(c_i,k_i) f_i(c_i,q_i) dc_i \, dk_i \tag*{(Integrating out $b_{-i}$)} \nonumber \\
&=  -\int_{\underline{k}_i}^{\bar{k}_i}\int_{\underline{c}_i}^{\bar{c}_i} \bigg(\rho_i(\bar{c}_i, k_i)  +  \int_{c_i}^{\bar{c}_i} N_i(z,k_i) dz\bigg) \, f_i(c_i,k_i)  dc_i \, dk_i \tag*{(As we need truthfulness)} \\
&=  -\int_{\underline{k}_i}^{\bar{k}_i}\int_{\underline{c}_i}^{\bar{c}_i} \rho_i(\bar{c}_i, k_i) f_i(c_i,k_i)  dc_i \, dk_i \nonumber \\ 
& \qquad - \int_{\underline{k}_i}^{\bar{k}_i}\int_{\underline{c}_i}^{\bar{c}_i} 
 N_i(z,k_i) dz \int_{\underline{c}_i}^{z}  \, f_i(c_i|k_i)  dc_i \; f_i(k_i) dk_i \tag*{(Changing order of integration)}\nonumber \\  
 &=  -\int_{\underline{k}_i}^{\bar{k}_i}\int_{\underline{c}_i}^{\bar{c}_i} \rho_i(\bar{c}_i, k_i) f_i(c_i,k_i)  dc_i \, dk_i \nonumber \\ 
& \qquad - \int_{\underline{k}_i}^{\bar{k}_i}\int_{\underline{c}_i}^{\bar{c}_i} 
 N_i(z,k_i) F_i(z|k_i) dz  f_i(k_i) dk_i \nonumber \\  
 &=  -\int_{\underline{k}_i}^{\bar{k}_i}\int_{\underline{c}_i}^{\bar{c}_i} \rho_i(\bar{c}_i, k_i) f_i(c_i,k_i)  dc_i \, dk_i \nonumber \\ 
& \qquad - \int_{\underline{k}_i}^{\bar{k}_i}\int_{\underline{c}_i}^{\bar{c}_i} 
 N_i(z,k_i) \frac{F_i(c_i|k_i)}{f_i(c_i|k_i)}   f_i(c_i, k_i) dz \, dk_i \label{eq-inter}
\end{align}
}
The last step is obtained by relabeling the variable of integration and simplifying.

Here, $\rho_i(\bar{c}_i, k_i)$ denotes the utility of a $MU_i$ when its true type is $(\bar{c}_i, k_i)$. With this type profile, the auctioneer by paying $\bar{c}_i$ can ensure both IR and IC, hence we can set $\rho_i(\bar{c}_i, k_i) = 0, \forall k_i \in [\underline{k}_i,\bar{k}_i]$. Applying this in the above equation and simplifying we get that the objective function of auctioneer is same in form to Equation~(\ref{eq:opt}). Consider Equation~(\ref{eq-inter}) and set $\rho_i(\bar{c}_i, k_i) = 0$ and simplification yields Equation~(\ref{eqn:opt_payment}).
By construction, the mechanism is BIC and IR. By hypothesis, as the auctioneer's objective is maximized, the mechanism is optimal.
\end{pfof}

\end{document}